\documentclass[
  aps,prd,twocolumn,superscriptaddress,reprint,
  amsmath,amssymb
]{revtex4-2}
\usepackage{graphicx}
\usepackage[colorlinks]{hyperref}
\begin{document}
\title{Wavelets for power spectral density estimation of gravitational wave data}
\author{Jin-Bao Zhu}
\author{Chao-Wan-Zhen Wang}
\author{Guo-Qing Huang}
\affiliation{Department of Physics, Nanchang University, Nanchang, 330031, China}
\affiliation{Center for Relativistic Astrophysics and High Energy Physics,
  Nanchang University, Nanchang, 330031, China}
\author{Fu-Wen Shu}
\email{shufuwen@ncu.edu.cn}
\affiliation{Department of Physics, Nanchang University, Nanchang, 330031, China}
\affiliation{Center for Relativistic Astrophysics and High Energy Physics,
  Nanchang University, Nanchang, 330031, China}
\affiliation{Center for Gravitation and Cosmology, Yangzhou University,
  Yangzhou, 225009, China}
\begin{abstract}
Power spectral density (PSD) estimation is a critical step in gravitational wave
(GW) detectors data analysis. The Welch method is a typical non-parametric
spectral estimation approach that estimates the PSD of stationary noise by
averaging periodograms of several time segments, or by taking the median of
periodograms to adapt to non-stationary noise. In this work, we propose a
wavelet-based approach for fast PSD estimation of both stationary and
non-stationary noise. For stationary noise, we apply wavelet smoothing to the
periodogram, avoiding the segmentation step in the Welch method, and enabling
PSD estimates with high frequency resolution and low variance. The wavelet
smoothing PSD outperforms Welch PSD in matched filtering and parameter
estimation. For non-stationary noise, we estimate the PSD by taking the median
of wavelet packet coefficients in each frequency bin, which offers greater
robustness than the traditional median periodogram method. This work introduces
a new PSD estimation approach for GW data analysis and expands the application
of wavelet methods in this field.
\end{abstract}
\maketitle
\section{Introduction}
GW data from ground-based observatories are one-dimensional time series,
which can be viewed as instances of stochastic processes or random
signals \cite{GWPA,AOGW}. In time series analysis, the PSD is one of the most
important second-order statistics. Many GW data analysis methods operate in the
frequency domain, and the application of these methods relies on reliable
estimation of the PSD \cite{GWPA,AOGW,AGTL}. This paper explores methods for
estimating the PSD and applies them to estimate the PSD of GW detector noise
(GW data).

GW data analysis has two main aspects: signal detection and parameter
estimation \cite{AGTL}. Signal detection aims to identify hidden GW signals
within the data. These signals are so weak that matched filter \cite{GFRF,FAAF}
becomes necessary to detect their presence. However, the performance of matched
filters depends on the accuracy of the PSD estimation. Parameter estimation,
on the other hand, seeks to determine the astrophysical source parameters
associated with the detected signals. The core of parameter estimation lies in
the log-likelihood \cite{AGTL}, which can be computed either in the time domain
(relying on the autocovariance function, ACF) or in the frequency domain
(relying on the PSD) \cite{ABHR}. Notably, the ACF and PSD are related via
the Wiener-Khinchin theorem \cite{GWPA,AOGW,ABHR}. In essence, both matched
filters and parameter estimators depend on the PSD because they require
a whitened noise background to function effectively. Since the PSD directly
impacts both matched filters and parameter estimators, we naturally want to
estimate the detector noise's PSD as accurately as possible.

PSD estimation methods generally fall into two classes: non-parametric and
parametric approaches \cite{SAOS,DSPP}. The typical non-parametric method is
the Welch method \cite{welch}, which divides the target time series data into
many overlapping segments of equal length, then takes the average of all the
segment periodograms. The Welch method works well for stationary noise but
cannot estimate PSD for non-stationary noise. If we change the averaging
operation in the Welch method to taking the median instead, this modified
Welch method can then estimate PSD for non-stationary noise. Many works
\cite{PEFCBC,PEFC,TTFL} use this approach for PSD estimation, and software
packages like Bilby \cite{bilby} and PyCBC \cite{pycbc} provide relevant
interfaces for it. The parametric approach involves building mathematical models
of the main noise sources, using AR(MA) models for fitting \cite{WONN,OPSI,ARSO}
or Bayesian inference methods \cite{BIFS,BPSE,BAWA,NSEM,PEGW} to obtain PSD
estimates. The main issue with non-parametric methods is that the PSD estimates
have large variance, requiring a trade-off between reducing variance and
losing frequency resolution. The main problem with parametric methods is their
dependence on models---if the noise modeling is inaccurate, then the PSD
estimation will also be inaccurate \cite{SAOS}.

Wavelet analysis provides an approach for noise PSD estimation. For the
stationary noise case, traditional non-parametric PSD estimation methods face a
fundamental trade-off: reducing PSD variance inevitably sacrifices frequency
resolution. Wavelet methods effectively overcome this limitation. In subsequent
sections, we empoly wavelet smoothing techniques to process periodograms,
yielding \emph{wavelet smoothing PSD} estimates. This approach can not only
effectively reduces PSD variance, but also does not require data segmentation,
thus avoiding the reduction of PSD frequency resolution. In addition, the
wavelet smoothing PSD not only perfectly inherits the advantages of
non-parametric methods in terms of computational efficiency, but also performs
better than Welch PSD in terms of matched-filtering and parameter estimation.
For the non-stationary noise scenario, PSD estimation methods based on the
averaging periodogram become invalid. A valid alternative is to first transform
the data into time-frequency domain using wavelet packet transform, then compute
median values across frequency bands, yielding what we call \emph{median wavelet
packet PSD}, in a manner analogous to median periodogram methods
\cite{PEFCBC,PEFC}. Especially the \emph{median Wilson-Daubechies-Meyer (WDM)
PSD} based on the WDM transform \cite{TFAG,TAWF} can not only adapt to the
non-stationary of the noise, but also is more robust than the median periodogram
PSD.

This paper is organized as follows. Section~\ref{sec:wavelets} describes the
wavelet methods employed in our work, including the continuous wavelet transform
(CWT), the discrete wavelet transform (DWT), and the wavelet packet transform
(WPT). Section~\ref{sec:stationary} addresses stationary noise scenarios, where
we introduce wavelet smoothing techniques to obtain non-parameter PSD estimates
that simultaneously achieve both high frequency resolution and low variance.
For non-stationary noise conditions, Section~\ref{sec:nonstationary} then
develops the PSD estimation approach based on WPT. Finally, conclusions are
provided in Section~\ref{sec:conclusions}.

\section{Wavelet methods}\label{sec:wavelets}
For a given finite-energy signal $f(t)$, its CWT is defined in the form of an
inner product as \cite{TLOW,AWTO}
\begin{equation}\label{eq:cwt}
  W(a,b) = \int_{-\infty}^{\infty}f(t)\psi_{a,b}^*(t)\,\mathrm{d}t.
\end{equation}
Here, $a,b\in\mathbb{R}$ and are called the scale parameter and translation
parameter respectively, with $a>0$ typically. The notation $\psi_{a,b}(t)=
\frac{1}{\sqrt{a}}\psi\bigl(\frac{t-b}{a}\bigr)$, where $\psi(t)$ is the wavelet
basis function (also known as the ``mother wavelet'').

The image $W(a,b)$ of an one-dimensional signal $f(t)$ under the CWT is the
scalogram representation of $f(t)$. The wavelet basis function $\psi(t)$ has
a centeral frequency $f_c$, which implies that $\psi(t)$ corresponds to
different frequencies $f$ at different scales $a$, with the scale-frequency
relationship given by
\begin{equation}
  a = \frac{f_c}{f}.
\end{equation}
Based on this correspondence, the scalogram can be interpreted as a spectrogram
representation. In other words, the CWT $W(a,b)$ of $f(t)$ also serves as a
time-frequency representation of the signal, allowing us to extract frequency
information at different time points from $W(a,b)$.

When implementing the CWT on computers, it is impossible to compute all possible
values of $a$ and $b$, so they must be discretized, thus converting the CWT into
a DWT. The scale parameter $a$ can be discretized using octave scales
\cite{APGT} or log-uniform scales \cite{NWTF}, while the discretization of the
translation parameter $b$ depends on $a$. Octave scales provides high frequency
resolution at low frequencies but low frequency resolution at high frequencies,
whereas log-uniform scales provides uniform frequency resolution. If signal
reconstruction (inverse transform) is not required, any suitable set of discrete
scales can be employed.

For some very special choices of $\psi(t)$ and $a,b$, there exists
$\psi_{j,k}(t)=2^{-j/2}\psi(2^jt-k)$, where $j,k\in\mathbb{Z}$, which can form
an orthogonal basis for $L^2(\mathbb{R})$ \cite{TLOW}. In this case, based on
multiresolution analysis theory, Eq.~\eqref{eq:cwt} has a fast orthogonal
algorithm, known as the Mallat decomposition \cite{AWTO}:
\begin{subequations}\label{eq:mallat-de}
\begin{align}
  A_{j-1}[k] &= \sum_lh^*[l-2k]A_j[l],\\
  D_{j-1}[k] &= \sum_lg^*[l-2k]A_j[l].
\end{align}
\end{subequations}
In these equations, the wavelet coefficients $A$ and $D$ are called
approximation coefficients and detail coefficients respectively, with their
subscripts corresponding to scales. $A_{j-1}$ and $D_{j-1}$ represent the
low-frequency and high-frequency components of $A_j$ respectively. $h$ and $g$
are the low-pass and high-pass filter coefficients, with $g[k]=(-1)^kh[1-k]$.
In practice, by taking $A_j$ as the input signal and selecting a specific
wavelet, the iterative process can be initiated. All wavelet coefficients
arranged hierarchically together form a time-frequency representation of the
input signal. This time-frequency representation is complete and perfectly
non-redundant, but the time-frequency tiling grid is non-uniform.

After performing the DWT, the wavelet coefficients can be directly manipulated
to alter the input signal. This step is typically achieved through thresholding.

The process inverse to the Mallat decomposition algorithm,
i.e. Eq.~\eqref{eq:mallat-de}, is called the Mallat reconstruction
algorithm \cite{AWTO}:
\begin{equation}\label{eq:mallat-re}
  A_j[l] = \sum_kh[l-2k]A_{j-1}[k] + \sum_kg[l-2k]D_{j-1}[k]. 
\end{equation}
The wavelet coefficients $A$ and $D$ used for reconstruction are usually
thresholded.

\begin{figure*}[t]
\centering
\includegraphics{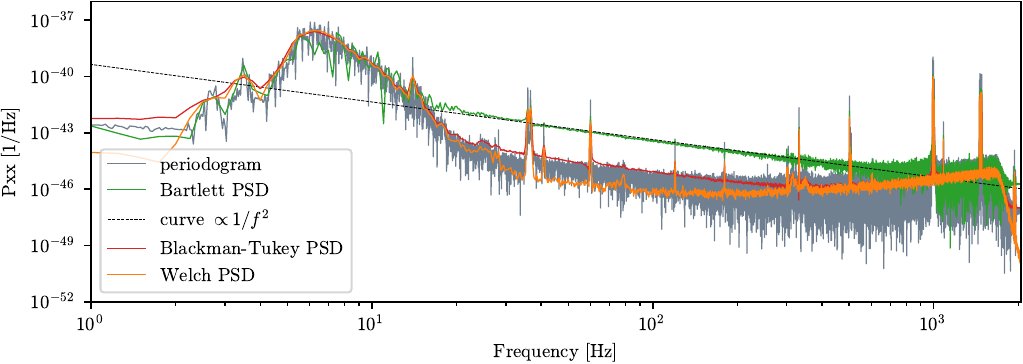}
\caption{One-side PSD estimates of GW150914 data obtained using different
methods. The gray curve indicates the periodogram with 65537 frequency points.
The green curve indicates the Bartlett PSD estimate using 4\,s data segments,
yielding 8193 frequency points. The red curve indicates the Blackman-Tukey
PSD estimate with a lag parameter of 8192, yielding 8193 frequency points.
The orange curve is the Welch PSD estimate, using 4\,s segments with 50\%
overlap and Blackman windowing, also yielding 8193 frequency points.
The black dashed line indicates the spectral leakage trend following $1/f^2$
scaling \cite{AGTL}. Notably, the periodogram, Bartlett PSD, and Blackman-Tukey
PSD all exhibit varying degrees of spectral leakage, with Bartlett being the
most severely affected.}
\label{fig:PSDs}
\end{figure*}

As shown in Eq.~\eqref{eq:mallat-de}, each decomposition step operates
only on the approximation coefficients $A$, which explains why the DWT produces
non-uniform time-frequency tiles. In contrast, the WPT decomposes both
approximation coefficients and detail coefficients. The wavelet packet
decomposition algorithm is given by formula \cite{AWTO}
\begin{subequations}\label{eq:wp-de}
\begin{align}
  D_{j-1}^{2n}  [k] &= \sum_lh^*[l-2k]D_j^n[l], \\
  D_{j-1}^{2n+1}[k] &= \sum_lg^*[l-2k]D_j^n[l].
\end{align}
\end{subequations}
In practice, one may take $D_j^n$ to be the input signal itself. After
performing the decomposition to a desired decomposition level, arranging
$D_{j-\mathrm{level}}^n$ by frequency yields a uniform time-frequency
representation of the input signal.

In addition to the aforementioned content, we will also employ the WDM
 transform. The technical details of WDM transform
can be found in References \cite{TFAG} and \cite{TAWF}. The WDM transform is
likewise a WPT, though its implementation differs from that described in
Eq.~\eqref{eq:wp-de}. Wavelet transform encompasses other variants as well,
including the stationary wavelet transform (SWT) and stationary wavelet packet
transform (SWPT) \cite{TSWT,LSWP,WMFT}. Regardless of the specific transform
type, however, they all represent particular discretization of the CWT image
$W(a,b)$.

\begin{figure*}[t]
\centering
\includegraphics{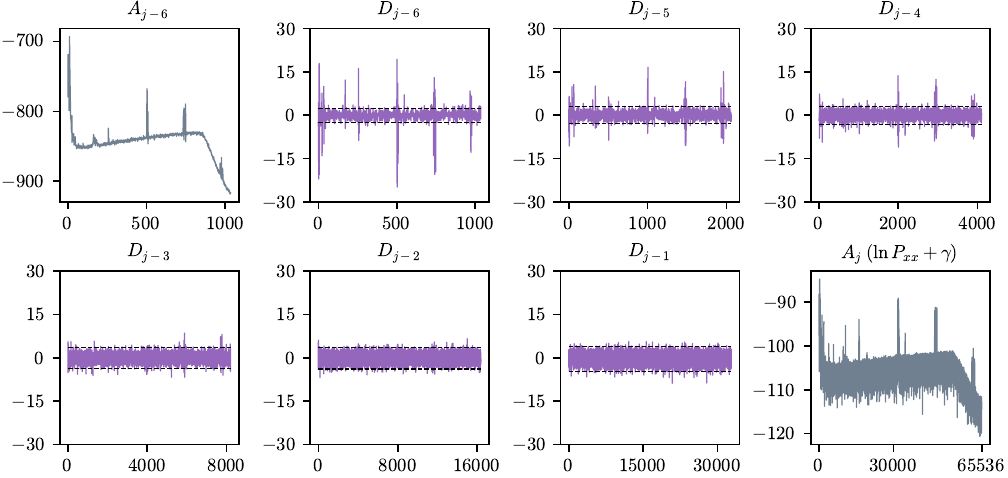}
\caption{The log-periodogram plus constant $\gamma$ ($A_j$) and its
decomposition results ($D_{j-1},\ldots,D_{j-6},A_{j-6}$) for GW150914 data
from H1 detector. We employed sym7 wavelet and Tukey window with $\alpha=0.08$.
The horizontal axis represents the index, while the black dashed horizontal
lines indicate thresholds calculated based on subsequent discussion. As can be
seen from the figure, $A_j$ contains significant fluctuations. After
decomposition, most of these fluctuations are distributed across the detail
coefficients at various levels. As the number of decomposition levels increases,
the wavelet coefficients become progressively coarser. The genuine components of
the log-PSD gradually become apparent in the detail coefficients (for example,
the sharp spectral lines visible in the detail coefficients), while the
approximation coefficient gradually reveals the overall trend of $A_j$.}
\label{fig:PSD-decomposition}
\end{figure*}

\section{Stationary noise}\label{sec:stationary}
\subsection{Averaging modified periodograms}
For a zero-mean stationary time series process $x[n]$, $n=0,1,\ldots,N-1$,
the periodogram is defined as \cite{SAOS,DSPP}
\begin{equation}\label{eq:cperiodogram}
  P_{xx}(f) = \frac{1}{N}\mathopen{}\mathclose{
    \left|\sum_{n=0}^{N-1}x[n]\mathrm{e}^{-2\pi\mathrm{i}fn}\right|^2}
    = \frac{1}{N}\bigl|\hat{x}(f)\bigr|^2,
\end{equation}
where $\hat{x}$ denotes the Fourier transform of $x$. In practice, we need to
discretize the above expression as
\begin{gather}
  P_{xx}[k] = \frac{1}{N}\mathopen{}\mathclose{
    \left|\sum_{n=0}^{N-1}x[n]\mathrm{e}^{-2\pi\mathrm{i}nk/N}\right|^2}
    = \frac{1}{N}\bigl|\hat{x}[k]\bigr|^2, \label{eq:dperiodogram}\\
    k = 0,1,\ldots,N-1. \nonumber
\end{gather}
The Fourier transform captures the frequency characteristics of a signal,
hence the periodogram reflects the signal's frequency distribution. In fact,
the periodogram serves as an estimate of the signal's PSD, though this estimate
is biased.  The periodogram exhibits high variance, fluctuating around the true
PSD \cite{SAOS}. To reduce these fluctuations, the periodogram must be smoothed.

An effective approach for smoothing the periodogram is Welch method of averaging
modified periodograms \cite{welch,SAOS,DSPP}, which proceeds in three steps:
\begin{itemize}
  \item[(1)] divide the $N$-point time series into overlapping segments
             of equal length;
  \item[(2)] apply a window to each segment and compute its periodogram with
             Eq.~\eqref{eq:dperiodogram};
  \item[(3)] average the periodograms of all segments to obtain the
             Welch PSD estimate.
\end{itemize}
The windowing operation in step (2) suppresses spectral leakage introduced by
the fast Fourier transform (FFT), while the averaging in step (3) reduces the
variance of the final PSD estimate. However, this comes at the cost of reduced
frequency resolution due to the segmentation in step (1). As the segment length
becomes shorter than the original time series, the resulting periodogram
contains fewer points (see Eq.~\eqref{eq:dperiodogram}). Although the
frequency range remains unchanged, the frequency resolution is consequently
lower.

Welch method cannot reduce variance while preserving frequency resolution,
necessitating a trade-off between frequency resolution and variance. Other
approaches, such as Bartlett method and the Blackman-Tukey method, similarly
require sacrificing frequency resolution to achieve reduced variance in PSD
estimation \cite{SAOS,DSPP}. Moreover, beyond this fundamental resolution-%
variance trade-off, both Bartlett and Blackman-Tukey methods inherently exhibit
more significant estimation errors. Fig.~\ref{fig:PSDs} displays one-side PSD
estimates obtained using different methods for the GW150914 data from the H1
detector. The dataset has a duration of 32\,s, a sampling rate of 4096\,Hz, and
consists of $N=131072$ points. In the discussion presented in
Section~\ref{sec:stationary}, all analyses are based on this dataset.

We aim to obtain a PSD estimate with both high frequency resolution and low
variance, which precludes data segmentation as used in Welch's method. Without
segmentation, however, the periodogram exhibits significant variance, shifting
the problem to finding an appropriate periodogram smoothing approach.
Fortunately, the DWT combined with thresholding provides an effective solution
to this problem \cite{WTTF,EEOP,SMBW}. In this paper, we refer to this method as
\emph{wavelet smoothing}. With wavelet smoothing, we avoid data segmentation
while achieving adjustable degrees of periodogram smoothing (though it should
be noted that excessive smoothing may obliterate spectral features).

\subsection{Wavelet smoothing PSD}\label{subsec:wtpsd}
As previously established, the periodogram of a stationary time series process
is given by Eq.~\eqref{eq:cperiodogram}. The periodogram is symmetric about zero
frequency, with its frequency range being $-f_s/2\leq f\leq f_s/2$, $f_s$, where
$f_s$ is the sampling rate of the time series. The periodogram has an important
property \cite{TSDAAT,WMFT}. When the second moment is finite and as
$N\to\infty$, $P_{xx}(f)$ follows the model
\begin{equation}
  P_{xx}(f) = \left\{\!\!\begin{array}{ll}
    S_{xx}(f)\chi_2^2/2, & 0<|f|<f_s, \\[1ex]
    S_{xx}(f)\chi_1^2,   & |f|=0~\text{or}~f_s. \\
  \end{array}\right.
\end{equation}
Here, $\chi_\eta^2$ denotes a $\chi^2$ distributed random variable with $\eta$
degrees of freedom, and $S_{xx}(f)$ represents the true PSD of the time series.
Utilizing properties of the Gamma distribution, one can derive \cite{TSAO,WMFT}
\begin{equation}\label{eq:PS-rel}
  \ln P_{xx}(f) + \gamma = \ln S_{xx}(f) + \epsilon(f),
\end{equation}
where $\gamma\approx0.57721$ is the Euler-Mascheroni constant, and $\epsilon(f)$
is a random variable with variance $\sigma_\epsilon^2=\pi^2/6$. Furthermore,
it can be shown that the probability density function of $\epsilon(f)$ follows
the Gumbel distribution
\begin{equation}\label{eq:gumbel}
  p_\epsilon(y) = \mathrm{e}^{y-\gamma-\mathrm{e}^{y-\gamma}}.
\end{equation}

Eq.~\eqref{eq:PS-rel} reveals that under logarithmic transformation, the
log-periodogram plus a known constant equals the log-PSD plus fluctuations with
zero mean and variance of $\pi^2/6$. If we could eliminate the $\epsilon(f)$
term in Eq.~\eqref{eq:PS-rel}, it would mean we have obtained a PSD
estimate from the periodogram. Below, we will explain how to use wavelet
smoothing method to obtain the PSD estimate from the left-hand side quantities
in Eq.~\eqref{eq:PS-rel}.

Letting $A_j=\ln P_{xx}+\gamma$, we obtain the following wavelet coefficients
according to Eq.~\eqref{eq:mallat-de}:
\[
  D_{j-1},D_{j-2},\ldots,D_{j-m};\quad A_{j-m}.
\]
where the positive integer $m$ represents the number of iterations (i.e.,
decomposition levels) in Eq.~\eqref{eq:mallat-de}. In practice, it is
unnecessary to decompose $A_j$ to the final level, so $m$ is typically small.
In Eq.~\eqref{eq:mallat-de}, the number of points in $A_{j-1}$ and
$D_{j-1}$ is half that of $A_j$, meaning the wavelet coefficients become
increasingly coarse with each iteration. The detail coefficients
$D_{j-1},D_{j-2},\ldots,D_{j-m}$ capture progressively coarser fluctuations in
the log-periodogram, while the approximation coefficient $A_{j-m}$ reflects its
increasingly coarse trend. Fig.~\ref{fig:PSD-decomposition} shows the
decomposition results of the log-periodogram of GW150914 data plus $\gamma$.
Note that to suppress spectral leakage, the data must be windowed before
computing the periodogram. From the Eq.~\eqref{eq:PS-rel}, the fluctuation
term $\epsilon(f)$ is distributed across the approximation coefficient and the
detail coefficients of every level.

\begin{figure}[t]
\centering
\includegraphics{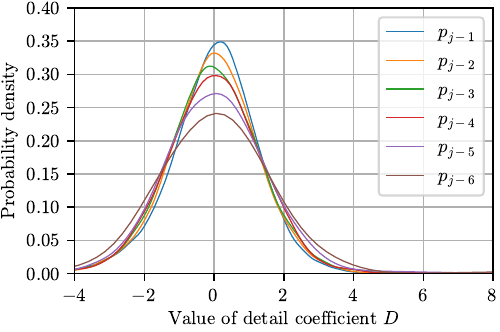}
\caption{Probability density distribution of detail coefficients of every level
in Fig.~\ref{fig:PSD-decomposition}.}
\label{fig:distribution}
\end{figure}

Ideally, the fluctuation $\epsilon(f)$ would be completely decomposed into the
detail coefficients at each level, while the approximation coefficient
$A_{j-m}$ would fully represent the log-PSD. In this case, one could simply set
all the detail coefficients to zero and then reconstruct the true log-PSD using
Eq.~\eqref{eq:mallat-re}. However, in reality, the detail coefficients
contain not only the fluctuation $\epsilon(f)$ but also parts of the log-PSD,
and the same is true for the approximation coefficients. This can also be seen
in Fig.~\ref{fig:PSD-decomposition}. To accurately reconstruct the true
log-PSD, we need to eliminate the fluctuation component $\epsilon(f)$ from the
detail coefficients $D$ as much as possible. To do this, we need to understand
the characteristics of the detail coefficients.

For many different types of data, the distribution of their detail coefficients
closely approximates the so-called generalized Gaussian distribution (GGD)
\cite{SMBW,ATFM}:
\begin{equation}\label{eq:GGD}
  p_{\mathrm{GGD}}(y) = \frac{\beta}{2\alpha\Gamma(1/\beta)}
    \exp[-(|y|/\alpha)^\beta],\quad \alpha,\beta>0.
\end{equation}
In our problem, $A_j$ was decomposed into 6 levels, with the distributions of
detail coefficients $D_{j-1},\ldots,D_{j-6}$ at each level shown in Fig.~%
\ref{fig:distribution}. As can be observed from the figure, these detail
coefficient distributions generally conform to the model \eqref{eq:GGD},
even though the probability density distribution of $\epsilon(f)$ should
theoretically follow Eq.~\eqref{eq:gumbel}.

Based on the model in Eq.~\eqref{eq:GGD}, we can apply percentile soft
thresholding \cite{ISAB,SMBW} to the detail coefficients to remove the
$\epsilon(f)$ component. The tail probability
\begin{equation}
  P(y) = \left\{\rule{0pt}{5.2ex}\!\!\begin{array}{ll}
    \int_{y}^\infty p(y)\,\mathrm{d}y,    & y\geq0, \\[1.2ex]
    \int_{-\infty}^{y} p(y)\,\mathrm{d}y, & y<0
  \end{array}\right.
\end{equation}
is related to the significance level $s$ and the threshold $\lambda$, with the
relationship given by \cite{WTTF}
\begin{equation}\label{eq:alpha-lambda}
  P(\lambda^-) = P(\lambda^+) = \frac{s}{2}.
\end{equation}
Eq.~\eqref{eq:alpha-lambda} is used to compute the upper thresholds $\lambda^+$
and lower thresholds $\lambda^-$. Since the detail coefficient distribution is
not symmetric about $y=0$, the upper and lower thresholds are also asymmetric.
The thresholds represented by the black dashed lines in Fig.~%
\ref{fig:PSD-decomposition} are listed in Table~\ref{tab:threshold}; these were
computed using Eq.~\eqref{eq:alpha-lambda}. Note that the choice of significance
level $s$ affects the resulting variance of $\epsilon(f)$, and $s$ should be
chosen such that the variance of $\epsilon(f)$ is close to $\pi^2/6$.

\begin{table}[t]
\begin{ruledtabular}
\begin{tabular}{lcc}
            & lower threshold $\lambda^-$ & upper threshold $\lambda^+$ \\
  \hline
  $D_{j-1}$ & $-4.85$ & $3.86$ \\
  $D_{j-2}$ & $-3.92$ & $3.60$ \\
  $D_{j-3}$ & $-3.68$ & $3.50$ \\
  $D_{j-4}$ & $-3.28$ & $3.10$ \\
  $D_{j-5}$ & $-2.77$ & $2.92$ \\
  $D_{j-6}$ & $-2.43$ & $2.33$
\end{tabular}
\end{ruledtabular}
\caption{The list of thresholds represented by the black dashed lines in
Fig.~\ref{fig:PSD-decomposition}, significance level $s=0.12$.}
\label{tab:threshold}
\end{table}

\begin{figure*}[t]
\centering
\includegraphics{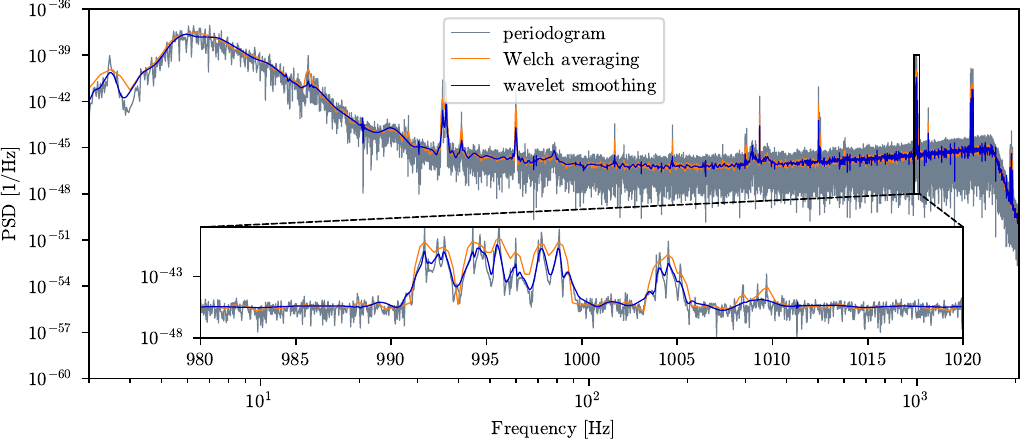}
\caption{Comparison between Welch PSD and wavelet smoothing PSD estimates.
The lower part of the figure shows a zoomed-in view of the 980--1020\,Hz
frequency range. The fluctuations in the original periodogram have been
effectively suppressed. While less visually prominent, the wavelet smoothing
PSD demonstrates higher frequency resolution and lower variance compared to
the Welch PSD.}
\label{fig:PSD-comparison}
\end{figure*}

We can now remove the $\epsilon(f)$ component from the detail coefficients $D$
by applying soft thresholding. After thresholding, we obtain the modified
coefficients $D_{j-1}^{\mathrm{th}},\ldots,D_{j-6}^{\mathrm{th}}$, which along
with the approximation coefficient $A_{j-6}$ can be substituted into Eq.~%
\eqref{eq:mallat-re} to reconstruct the final PSD estimate. Fig.~%
\ref{fig:PSD-comparison} compares the Welch PSD estimate with the wavelet
smoothing PSD estimate.

\begin{table*}[t]
\begin{ruledtabular}
\begin{tabular}{lccc}
  Method &
  Frequency resolution &
  Quality factor $Q$ &
  Max matched-filter SNR \\
  \hline
  Welch averaging   & 1/4\,Hz,  $N=8193$  & 0.001353 & 19.27 \\
  Wavelet smoothing & 1/32\,Hz, $N=65537$ & 0.001528 & 19.38
\end{tabular}
\end{ruledtabular}
\caption{Performance comparison between the Welch PSD and the wavelet smoothing
PSD shown in Fig.~\ref{fig:PSD-comparison}. The Welch method uses a Blackman
window, with a segment length of 4\,s and 50\% overlap. The wavelet smoothing
method uses a Tukey window with $\alpha=0.08$, sym7 wavelet, 6 levels of
decomposition, reflection extension mode, significance level $s=0.12$, and soft
thresholding processing.}
\label{tab:performance}
\end{table*}
\begin{table*}[t]
\begin{ruledtabular}
\begin{tabular}{lcccc}
  Wavelet &
  Significance level &
  $\mathrm{Var}[\epsilon(f)]$ &
  Quality factor $Q$ &
  Max matched-filter SNR \\
  \hline
  sym6    & 0.10 & 1.650 & 0.001285 & 19.73 \\
  sym6    & 0.11 & 1.646 & 0.001284 & 19.71 \\
  sym6    & 0.12 & 1.640 & 0.001288 & 19.68 \\
  sym7    & 0.11 & 1.646 & 0.001520 & 19.41 \\
  sym7    & 0.12 & 1.640 & 0.001528 & 19.38 \\
  sym7    & 0.25 & 1.586 & 0.001464 & 19.13 \\
  sym8    & 0.11 & 1.645 & 0.001318 & 19.72 \\
  sym8    & 0.50 & 1.501 & 0.001328 & 18.92 \\
  bior4.4 & 0.11 & 1.646 & 0.001210 & 19.64 \\
  bior4.4 & 0.50 & 1.497 & 0.001228 & 18.83 \\
  db7     & 0.13 & 1.643 & 0.001506 & 19.64 \\
  db1     & 0.15 & 1.645 & 0.001595 & 19.36
\end{tabular}
\end{ruledtabular}
\caption{Performance of the wavelet smoothing PSD under different wavelet basis
and significance levels. The significance level affects the variance of
$\epsilon(f)$, and a suitable significance level can be inferred based on this
variance. As shown in the table, if the wavelet type and significance level are
chosen appropriately, the wavelet smoothing PSD performs well. Otherwise, its
performance may even be worse than that of the Welch PSD.}
\label{tab:comparison}
\end{table*}
\begin{table*}[t]
\begin{ruledtabular}
\begin{tabular}{lccc}
  Method & Frequency resolution & Max network SNR & Bayesian factor \\
  \hline
  Welch averaging      & 1/4\,Hz   & 24.95 & 285.50 \\
  Wavelet smoothing    & 1/128\,Hz & 25.79 & 311.71
\end{tabular}
\end{ruledtabular}
\caption{Performance of Welch PSD and wavelet smoothing PSD in parameter
estimation.}
\label{tab:pe}
\end{table*}

\subsection{Performance}
The evaluation criteria for PSD estimation encompass multiple aspects, including
frequency resolution, quality factor \cite{DSPP}, matched-filter
signal-to-noise ratio (SNR) \cite{GFRF,FAAF}, among others. The frequency
resolution should be as high as possible to characterize fine details of the
PSD. The quality factor is defined as $Q=[\mathrm{E}(\mathrm{PSD})]^2/
\mathrm{Var}(\mathrm{PSD})$, that is, the square of the mean of the periodogram
divided by its variance. The quality factor reflects the stability and
smoothness of the PSD estimate and should be as large as possible. Furthermore,
the PSD directly impacts matched-filter performance. Namely, only PSD estimates
that accurately capture the background noise characteristics can enable the
matched filter to function effectively. The wavelet smoothing method is a
non-parametric approach to PSD estimation. To evaluate its performance, we will
compare it with the Welch method.

In the previous subsection, we obtained a PSD estimate for the GW150914 data
from H1 detector using the wavelet smoothing method, as shown by the blue curve
in Fig.~\ref{fig:PSD-comparison}. Table~\ref{tab:performance} lists the
frequency resolution, quality factor $Q$, and matched-filter performance of this
PSD estimate. The results demonstrate that the wavelet smoothing PSD achieves
higher frequency resolution, better quality factor, and improved matched-filter
SNR compared to the Welch PSD. Since the wavelet smoothing method does not
require data segmentation, it incurs no loss of frequency resolution. The
greater $Q$ of the wavelet smoothing PSD indicates it produces smoother
estimates than the Welch method. Furthermore, the superior matched-filter
performance demonstrates that the wavelet smoothing PSD more accurately captures
the characteristics of the background noise.

\begin{figure}[t]
\centering
\includegraphics{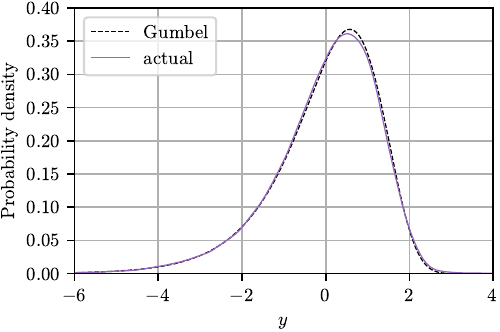}
\caption{The theoretical probability density distribution of $\epsilon(f)$
(Eq.~\eqref{eq:gumbel}, black curve) and its actual distribution
(purple curve).}
\label{fig:ef-distribution}
\end{figure}

According to Eq.~\eqref{eq:PS-rel}, the fluctuation term is given by
$\epsilon(f)=\ln P_{xx}(f)+\gamma-\mathrm{PSD}$. Theoretically, the distribution
of $\epsilon(f)$ should follow Eq.~\eqref{eq:gumbel}, and its actual
distribution is shown as the solid line in Fig.~\ref{fig:ef-distribution}.
The actual distribution of $\epsilon(f)$ agrees well with theoretical
predictions. The theoretical variance of $\epsilon(f)$ is $\sigma_\epsilon^2=
\pi^2/6$,  while in practice, it is $1.640\approx\pi^2/6=1.645$. This result can
be used to determine the appropriate significance level. The significance level
$s=0.12$ in Table~\ref{tab:threshold} was chosen to ensure that the variance of
$\epsilon(f)$ is close to $\pi^2/6$.

The performance of wavelet smoothing PSD estimation is influenced by multiple
factors: window function, wavelet basis, decomposition level, extension mode,
significance level, and thresholding method. In the present work, we fixed the
window function as a Tukey window with $\alpha=0.08$, maintained decomposition
level as 6, used reflect extension mode, and applied soft thresholding. Under
these conditions, Table~\ref{tab:comparison} presents the performance of the
wavelet smoothing PSD for different wavelet basis and significance levels.

In addition to the results in Table~\ref{tab:comparison}, we estimated the PSD
using both Welch method and wavelet smoothing method for 128\,s of off-source
data (GPS time 1126259332.4--1126259460.4\,s), followed by parameter estimation
with the Bilby \cite{bilby} software package. For the Welch method, a segment
length of 4\,s, 50\% overlap, and a Blackman window were used. The wavelet
smoothing method used the bior3.3 wavelet basis, with all other settings are
same as those listed in Table~\ref{tab:performance}. The final results are
listed in Table~\ref{tab:pe}.

\section{Non-stationary noise}\label{sec:nonstationary}
\begin{figure*}[t]
\centering
\includegraphics{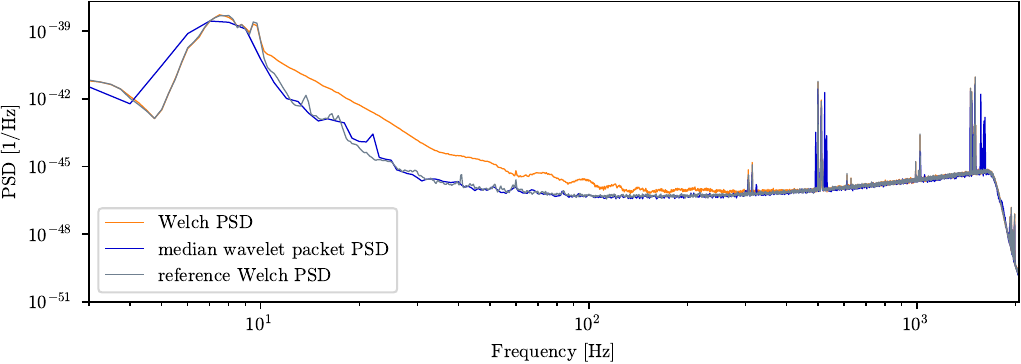}
\caption{PSD estimates using different methods. The orange and blue curves
represent PSD estimates based on 256\,s of L1 detector data (for GW170817) from
GPS time 1187008755--1187009011\,s, which contains a loud transient noise.
This transient occurred at approximately GPS time 1187008881.4\,s and can be
observed in Fig.~\ref{fig:GW170817-wpt} and the top and middle panels of
Fig.~\ref{fig:3fig}. The gray curve, labeled as the reference Welch PSD,
is based on 256\,s of quasi-stationary L1 data from GPS time
1187008625--1187008881\,s.}
\label{fig:wmpsd-robust}
\end{figure*}

In the previous section, we treated the GW150914 data as approximately
stationary noise. However, the noise in GW detectors is not
stationary. Its non-stationarity mainly includes long-duration adiabatic drifts
in the power spectrum and short-duration noise transients \cite{AGTL,TFAG}.
For non-stationary noise, global PSD estimation methods (such as Welch method)
based on the Fourier transform become invalid. This is because the Fourier
transform of time-domain data is a superposition of all frequency components,
and the presence of non-stationary noise leads to an inaccurate frequency-domain
description of the detector noise. In other words, the influence of
non-stationary noise will be reflected in the periodogram defined by
Eq.~\eqref{eq:cperiodogram} or Eq.~\eqref{eq:dperiodogram}, making
such periodograms ineffective representations of background noise. The orange
curve in Fig.~\ref{fig:wmpsd-robust} shows the Welch PSD estimate for L1
detector data in the GPS time 1187008755--1187009011\,s, which contains a very
loud transient noise. Due to the contribution of this transient, the Welch PSD
exhibits significantly elevated power in the 10--500\,Hz range compared to the
reference Welch PSD, clearly illustrating the impact of non-stationary noise on
PSD estimation.


To describe non-stationary noise, one can use the evolutionary power spectrum
(EPS) \cite{ESNP}. The goal of the EPS is to construct a time-frequency
distribution that captures the energy distribution of the noise across time and
frequency, along with its dynamic evolution. A common way to construct
time-frequency distributions is through the short-time Fourier transform,
but wavelet transforms provide a better alternative  \cite{AWTO}. Since the
wavelet transform is a time-frequency representation of one-dimensional data,
it naturally reflects the evolution of frequency over time. In fact, up to a
normalization factor, the EPS can be defined as the squared modulus of the
wavelet (or wavelet packet) transform \cite{APGT,WPAA,AWTO,LAEE}. The term
``wavelet transform'' here may refer to the continuous, discrete, or stationary
version. EPS accurately captures the variation of the signal's frequency content
over time and can be used for whitening, SNR computation, and likelihood
evaluation, among other applications in time-frequency domain \cite{TFAG}.

\begin{figure}[t]
\centering
\includegraphics{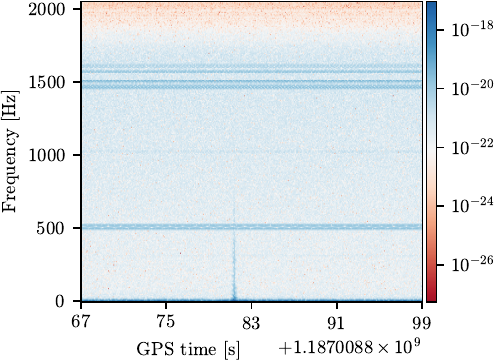}
\caption{The WPT results for GW170817 data from L1 detector
(\texttt{L-L1\_GWOSC\_4KHZ\_R1-1187008867-32.hdf5}). The transform was
implemented using the sym20 wavelet basis with 8 decomposition levels.
Increasing the decomposition level improves the frequency resolution. From the
figure, one can observe the signal's frequency evolution over time, as well as
the presence of a transient noise.}
\label{fig:GW170817-wpt}
\end{figure}

Similar to the median periodogram PSD method, one can also estimate the PSD of
non-stationary noise by taking the median of each frequency band in the EPS.
Due to the orthogonality of WPT, this approach offers high
computational efficiency, uniform frequency resolution, and flexible
time-frequency resolution selection, making it our preferred method for EPS
computation. Fig.~\ref{fig:GW170817-wpt} displays the WPT
results (using Eq.~\eqref{eq:wp-de}) for the GW170817 data from L1 detector.
By taking the squared modulus of the wavelet packet coefficients (wavelet packet
transform results), computing row-wise medians, and multiplying by a
normalization factor, we obtain a PSD estimate that mitigates the effects of
non-stationary noise. In Fig.~\ref{fig:wmpsd-robust}, the blue curve shows this
median wavelet packet PSD estimate (level=11, sym20 wavelet basis) for the same
GW170817 dataset. Except for a few narrow frequency bands, the result closely
matches the reference Welch PSD, demonstrating its effectiveness for
non-stationary noise environments.

Although the median wavelet packet PSD shown in Fig.~\ref{fig:wmpsd-robust}
can mitigate the effects of non-stationary noise, it is itself inaccurate. By
examining Figs.~\ref{fig:wmpsd-robust} and \ref{fig:GW170817-wpt}, we observe
spurious sharp spectral lines in the median wavelet packet PSD near 20\,Hz,
500\,Hz, and 1600\,Hz. We therefore conclude that the sym wavelet family is
not suitable for WPT of GW data. The WDM wavelet basis, with its
excellent time-frequency localization, demonstrates superior transient signal
processing capabilities and is better suited for this PSD estimation task
\cite{TAWF,TFAG}.

\begin{figure*}[t]
\centering
\includegraphics{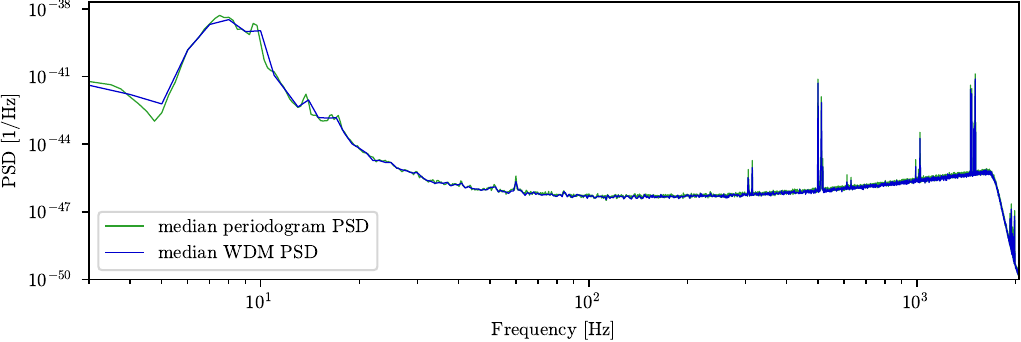}
\caption{Comparison between median periodogram PSD and median WDM PSD estimates.
The median periodogram PSD was computed using data segments of 4\,s with 87.5\%
overlap. Its frequency resolution is 1/4\,Hz, while the frequency resolution of
the median WDM PSD is 1\,Hz.}
\label{fig:median-PSD}
\end{figure*}

Using the same non-stationary data as in Fig.~\ref{fig:wmpsd-robust}, we
obtained a PSD estimate within the WDM framework (median WDM PSD), shown as the
blue curve in Fig.~\ref{fig:median-PSD}. For comparison, the green curve in
Fig.~\ref{fig:median-PSD} shows the result obtained using the median
periodogram method. While the median periodogram method can also resist the
influence of non-stationary noise, the figure shows that at sharp spectral lines
around 60\,Hz, 300\,Hz, and 500\,Hz, the median WDM PSD is weaker than the
median periodogram PSD, indicating that the median WDM PSD is more robust
\cite{TAWF}.

\begin{figure}[ht]
\centering
\includegraphics{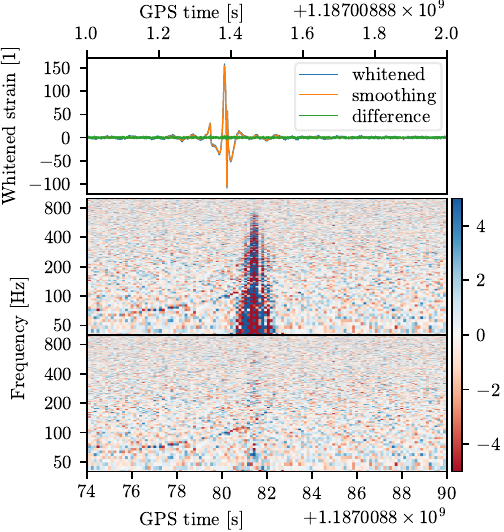}
\caption{The top panel displays a 1\,s segment of whitened strain (blue curve)
along with its smoothed version (orange curve), with the green curve showing
their pointwise difference. The middle panel presents the WDM transform of a
16\,s whitened strain segment, where the inspiral track of the GW signal is
visible in the lower left of panel. The bottom panel shows the WDM transform
of the green curve from the top panel. The transient noise has been largely
removed, revealing a more complete GW signal.}
\label{fig:3fig}
\end{figure}

Using the median WDM PSD, we whitened the GW data shown previously in
Fig.~\ref{fig:GW170817-wpt}. The whitened strain for a selected time segment
appears as the blue curve in the top panel of Fig.~\ref{fig:3fig},
where a loud transient noise is clearly visible. The middle panel of
Fig.~\ref{fig:3fig} displays the WDM transform of this whitened strain
(blue curve), with the inspiral track of the GW signal emerging in the lower
left region of the panel. The orange curve in the top panel of
Fig.~\ref{fig:3fig} is the result of applying the wavelet smoothing method
described in Section~\ref{subsec:wtpsd} to the blue curve. Using a soft
threshold designed according to the GW signal strength, we conclude this orange
curve primarily contains the transient noise component rather than the GW signal.
The green curve represents the difference of subtracting the orange curve from
the blue curve, and its WDM transform is shown in the bottom panel of Fig.~%
\ref{fig:3fig}. It can be seen that the transient noise has been largely
removed, and the portion of the GW signal previously obscured by the transient
has become visible.

\section{Conclusions}\label{sec:conclusions}
PSD estimation is a critical step in GW data analysis. There are many
traditional non-parametric PSD estimation methods for stationary noise,
but they require a trade-off between frequency resolution and variance,
making it impossible to achieve PSD estimates with both high frequency
resolution and low variance. In this work, we employ the wavelet smoothing
method for stationary noise, which achieves PSD estimates with both high
frequency resolution and low variance. This PSD estimation yields higher
matched-filter SNR, higher Bayes factors in parameter estimation, and better
frequency resolution compared to the Welch PSD estimate. The wavelet smoothing
PSD discussed here is influenced by multiple factors, but only the significance
level has a notable effect. When the variance of $\epsilon(f)$ is close to
$\pi^2/6$, the chosen significance level is considered appropriate.

For non-stationary noise, the median periodogram method can be employed for PSD
estimation, while the median wavelet packet method offers a new alternative.
Since the squared modulus of the wavelet transform can be treated as an
EPS, conventional WPT algorithms can be applied to map GW data into the
time-frequency domain, and then the median across each frequency band can be
taken to obtain the median wavelet packet PSD. However, due to limitations
inherent in conventional wavelets, this PSD estimate lacks sufficient accuracy.
The WDM wavelet, with its adaptability to transient signals, is suitable for
PSD estimation of non-stationary noise. PSD estimates based on the WDM transform
are more robust than those based on the median periodogram method \cite{TFAG}.

The PSD estimation methods discussed in this paper belong to the category of
non-parametric approaches and offer high computational efficiency. For the
wavelet smoothing method, the key computational steps consist of: one FFT,
one DWT, and one IDWT. For the median WDM method, the primary step is performing
a WDM transform, which itself can be decomposed into multiple FFTs. All these
main operations are orthogonal transforms, so there are no computational
efficiency concerns.

\section*{Acknowledgements}
This work is supported by the National Natural Science Foundation of China with
Grant No.~12375049, and Key Program of the Natural Science Foundation of
Jiangxi Province under Grant No.~20232ACB201008. DWT, IDWT, and WPT were
implemented using the \mbox{PyWavelets} \cite{PAPF} software package.
For the WDM transform, the \mbox{PyWavelet} \cite{TFAG,WDMT,WDMWT} library was
employed.
\bibliography{wtpsd.bib}
\end{document}